\documentclass[aps,reprint,prl,superscriptaddress]{revtex4-1}
\pdfoutput=1
\usepackage{graphicx}
\usepackage{amssymb,amsmath}
\usepackage{times}
\usepackage{subfigure}
\usepackage{color}
\usepackage{textcomp} 
\usepackage[bold]{hhtensor} 
\usepackage[colorlinks=true,linkcolor=blue,citecolor=red,urlcolor=black,pagebackref=true]{hyperref}
\newcommand{\sinn}{Si$_3$N$_4$\text{ }}
\newcommand{\sio}{SiO$_2$\text{ }}

\begin{document}
\title{Eliminating Structural Loss in Optomechanical Resonators Using Elastic Wave Interference}

\author{Mian Zhang}
\affiliation{School of Electrical and Computer Engineering, Cornell University, Ithaca, New York 14853, USA.}

\author{Gustavo Luiz}
\affiliation{Instituto de F\'{i}sica, Universidade Estadual de Campinas, 13083-970, Campinas, SP, Brazil.}

\author{Shreyas Shah}
\affiliation{School of Electrical and Computer Engineering, Cornell University, Ithaca, New York 14853, USA.}

\author{Gustavo Wiederhecker}
\affiliation{Instituto de F\'{i}sica, Universidade Estadual de Campinas, 13083-970, Campinas, SP, Brazil.}

\author{Michal Lipson$^{\ast}$}
\affiliation{School of Electrical and Computer Engineering, Cornell University, Ithaca, New York 14853, USA.}
\affiliation{Kavli Institute at Cornell for Nanoscale Science, Ithaca, New York 14853, USA.}
\date{\today}

\topmargin -1.5cm
\oddsidemargin -0.3in
\textwidth 18cm 
\textheight 23.2cm

\baselineskip12pt

\begin{abstract}
\begin{center}
\small{$^{\ast}$To whom correspondence should be addressed; e-mail:  ml292@cornell.edu}\\
\end{center}
Optomechanical resonators suffer from the dissipation of mechanical energy through the necessary anchors enabling the suspension of the structure. Here we show that such structural loss in an optomechnaical oscillator can be almost completely eliminated through the destructive interference of elastic waves using dual-disk resonators. We also present both analytical and numerical model that predicts the observed interference of elastic waves. Our experimental data reveal unstressed \sinn devices with mechanical Q-factors up to $10^4$ at mechanical frequencies of $f=102$ MHz ($fQ = 10^{12}$) at room temperature. 
\end{abstract}
\maketitle

\date{\today}
\newcommand{\nocontentsline}[3]{}
\newcommand{\tocless}[2]{\bgroup\let\addcontentsline=\nocontentsline#1{#2}\egroup}

Optomechanical resonators have fostered record detection of ultra-weak forces~\cite{Gavartin:2012el}, preparation of micromechanical oscillators close to their motional quantum ground states~\cite{ChaAleSaf1110,Verhagen:2012ei}, enabling self-sustaining mechanical oscillator dynamics \cite{Marquardt:2006kra,Poot:2012ta,Zhang:2012ks}, and optomechanical photo-detection~\cite{Tallur:2013hb}. But like all micromechanical resonators, their performance suffers from the dissipation of mechanical energy. The dissipation of mechanical energy in such devices reduces their sensitivity, shortens their coherence time, increases their power consumption and degrades the phase noise performance~\cite{Ekinci:2005hh,Vahala:2008wp}. This mechanical dissipation is often dominated by anchor losses at the necessary supporting clamps~\cite{Photiadis:2004kb,Lifshitz:2002ge,Cole:2011eh}, among other mechanisms responsible for the overall dissipation such as thermo-elastic damping~\cite{Lifshitz:2000dj}, phonon scattering~\cite{Lifshitz:2002ge}, and defect relaxations~\cite{Arcizet:2009ix}. Recent efforts in reducing anchor loss in micromechanical devices include using spoke design~\cite{Anetsberger:2008di,Cole:2011eh}, phononic bandgaps~\cite{Hsu:2011gf,Eichenfield:2009wl}  and materials with high internal stress~\cite{Verbridge:2006ix}. The spoke design creates an artificial bottleneck of energy flow at the cost of structural rigidity, whereas phononic bandgap materials are less suitable for lower frequency resonators as the size of the unit cell scales up and they occupy larger real estate.
\begin{figure}[htbp]
\begin{center}
\includegraphics[scale=1]{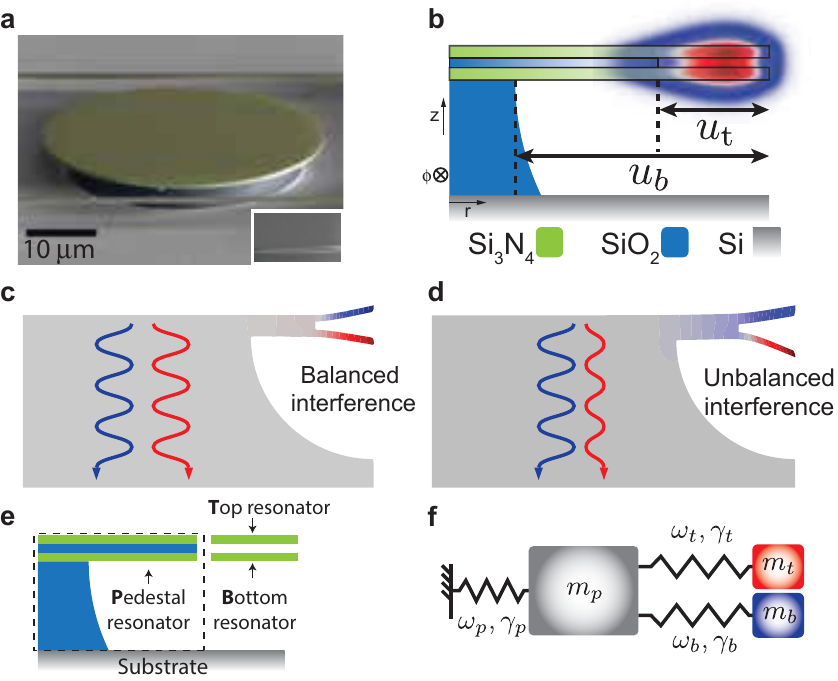}
\caption{\label{fig:intro} \textbf{Device schematic}. (a) Scanning electron micrograph of the fabricated device. The inset is a close-up of the freestanding double-disk edges. The two horizontal strings are for supporting tapered fibers. (b) Schematic of the cross section of the device, $u_t$ and $u_b$ are the undercut depth of the top and bottom layers, respectively. The false-color scale shows the transverse electric optical mode profile which spans the top and the bottom disks. (c,d) Finite-element model showing the impact of the thickness difference of the top and bottom cantilever, leading to an unbalanced interference of the elastic wave emitted by the moving edges. (e, f) A lumped theoretical model consists of three masses: $m_t$ and $m_b$ for the two edges and $m_p$ for the pedestal, each with mechanical frequencies $\omega_t$, $\omega_b$, $\omega_p$ and damping rate $\gamma_t$, $\gamma_b$, $\gamma_p$}.
\end{center}
\end{figure}

\begin{figure*}[htbp]
\begin{center}
\includegraphics[scale=1]{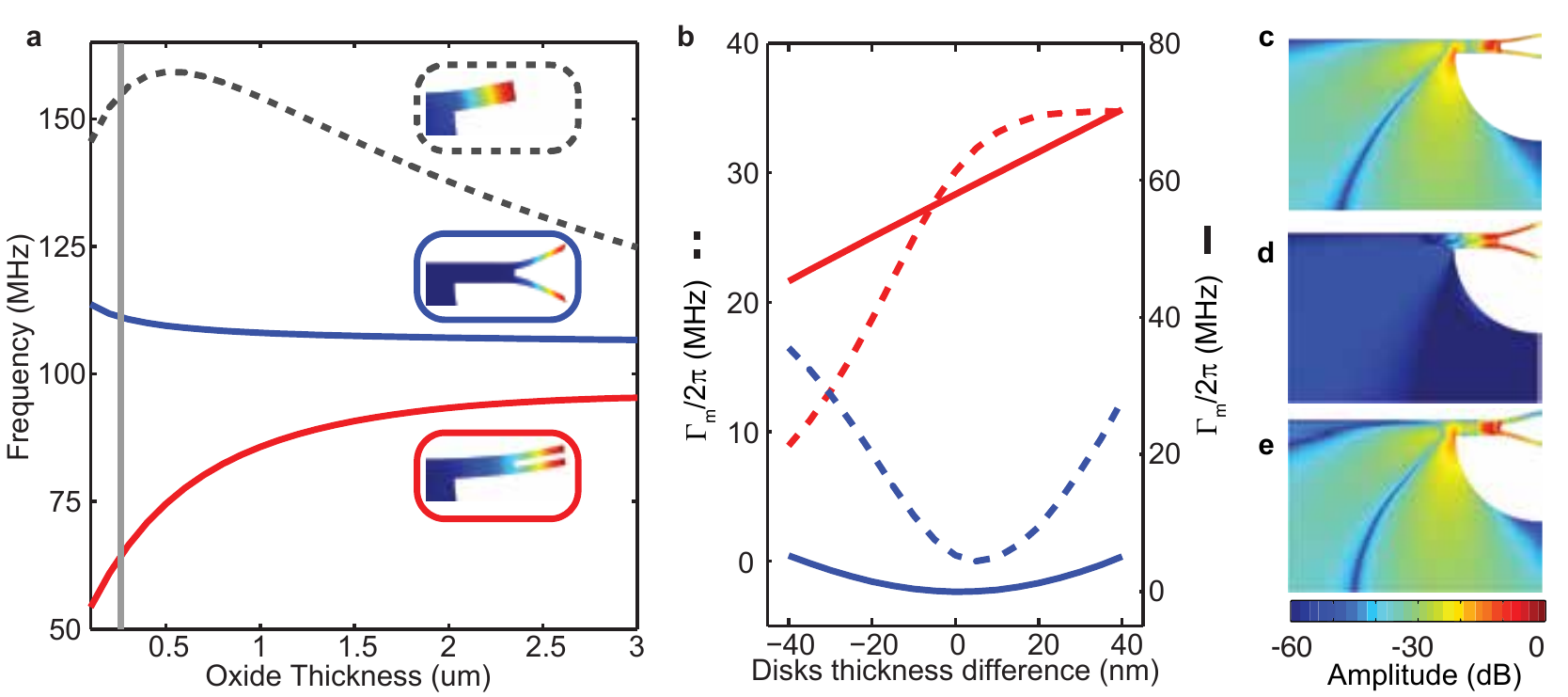}
\caption{\label{fig:model} \textbf{Device Simulations}. (a) Dispersion of mechanical frequencies as function of middle SiO$_2$ thickness; grey-dashed: $p$ mode, solid-blue: AS mode, solid-red: S mode, vertical solid-grey shows position of 200nm SiO$_2$ thickness.  (b) Attenuation as function of top and bottom disks thicknesses difference for 3$\mu$m (dashed, left scale) and 200~nm (solid, right scale) middle SiO$_2$ thickness, for the AS (blue) and S (red) modes. (c-e) $\hat{z}$ component of mechanical Poynting vector spatial distribution (false-color scale) for a top disk thicker  $\delta t=20$ nm (c), equal $\delta t=0$ (e), and thinner $\delta t=-20$ nm (f) than the bottom one.}
\end{center}
\end{figure*}

Here we show that the structural loss of an optomechanical oscillator can be effectively eliminated through the destructive interference of elastic waves, by emulating the principle of a tuning fork resonator. A tuning fork resonator produces a long lasting sound when excited, as a result of its high mechanical quality factor. When a tuning fork vibrates, its two prongs oscillate 180 degrees out of phase. The elastic wave produced from each prong largely cancels out leading to no net motion therefore no loss at the base. Here, in order to create the tuning fork effect, we use double-disk optomechanical resonators~\cite{Wiederhecker:2009ex,Rosenberg:2009vc}. 
     
We emulate the tuning fork principle using a dual-disk resonator, consisting of a pair of thin silicon nitride (\sinn) disks separated by a narrow gap (Fig. 1a,b). The thin \sio sacrificial layer mechanically couples the top and bottom resonator, allowing the mechanical waves to interfere. This sub-wavelength gap also results in the evanescent coupling of the optical fields, creating coupled optical modes that span both the top and the bottom disks. The attractive optical forces efficiently excite the antisymmetric mechanical modes, illustrated in Fig. 1c,d as the freestanding edges moves in opposite directions. The symmetric mechanical mode is however much less sensitive to the optical excitation. 
     
When the two freestanding edges are identical, the structural dissipation of the antisymmetric mechanical mode is minimized. The origin of the anchor loss in our structures is due to the displacement induced in the clamping area by the oscillation of each freestanding edge. Such displacement radiates elastic waves towards the pedestal and the substrate and therefore dissipates energy from the mechanical resonance (Fig.~1c,d). The antisymmetric mechanical mode excited experiences much less structural loss than the symmetric mechanical mode due to the destructive interference between the elastic wave radiated from the top and the bottom disks. 
     
In order to gain a physical intuition of the dissipation process, the dual-disk mechanical mode structure could be dissembled into simpler building blocks, the two freestanding edges (resonators T and B) emulating a tuning fork and the pedestal (resonator P), as depicted in Fig. 1e. Since all the mechanical energy inside the pedestal leaks to the bulk substrate, the structural loss rate of the resonator can therefore be established as the mechanical coupling rate between the freestanding resonators and the pedestal resonators. The higher the coupling between the freestanding edges and the pedestal resonator the more energy dissipation there is through the pedestal. 
     
The coupling strength of the freestanding edges to the pedestal is reflected in their dispersion curves as a function of the midlle \sio layer thickness ($t_m$).  In the case where there is no coupling, the mechanical frequencies of the edge modes would be independent of $t_m$, which is a parameter of the pedestal mode. Therefore the more sensitive the mode frequencies are to the \sio layer thickness $t_m$, the stronger the coupling is to the pedestal. We numerically investigate the coupling strength between the freestanding edge and the pedestal resonators using a finite-element (FEM) solver (COMSOL) through varying the thickness $t_m$ of the middle \sio layer . As shown in Fig. 2a, the antisymmetric (blue curve) mechanical mode stiffens and the symmetric (red curve) edge mode softens as $t_m$ reduces. The rapid softening of the symmetric mechanical mode indicates a strong mechanical coupling to the pedestal mode. Whereas the antisymmetric mode displays an almost flat dispersion relation to $t_m$, indicating that the antisymmetric mode is insensitive to $t_m$ of the pedestal resonator and therefore is weakly coupled to the pedestal mode. This is due to the canceling of the elastic wave from the two counter oscillating edges. 
 
The vital role of the thickness difference in the freestanding edges can be visualized through the mechanical dissipation rate shown in Fig. 2b. Our numerical simulation of the structural loss rate as a function of the thickness difference between the two edges in Fig. 2b confirms that indeed the loss rate is a minimum when the two disks are of equal thickness. Note that for a thick sacrificial layer (dashed curves) the minimum structural loss for the antisymmetric edge mode occurs when the top disk is slightly thicker. This is due to the symmetry breaking from the finite undercut radius of the bottom disk. Figs. 2c-e show the z-component of the mechanical energy flow (mechanical Poynting vector) for three top disk thickness differences when $t_m = 200$~nm. It is clear that the elastic wave radiation into the pedestal is drastically reduced when the two disks are of equal thickness.

We develop a pre-compensation technique to fabricate the freestanding edges of the double-disk structure and ensure that they are equal in thickness. We deposited a 240/200/220 nm  \sinn /\sio/\sinn film stack on a silicon wafer with 3 $\mu$m of thermal \sio. The stoichiometric \sinn films are deposited via low pressure chemical vapor deposition technique and the \sio layer is deposited via plasma-enhanced chemical vapor deposition and subsequently N$_2$ annealed at 1100 $^{\circ}$C over 1 hour. The 20 nm difference in the thickness of the two \sinn layers is designed to pre-compensate the change in their relative thickness as a result of the releasing wet etching step. We pattern the wafer with e-beam lithography and transfer the pattern with reactive ion etching (CHF$_3$/O$_2$). The devices are then undercut in a buffered oxide etch (6:1). This wet etching process has a finite selectivity to \sinn and \sio, roughly 1:100. Therefore it not only etches \sio at 80 nm/min but also removes \sinn at a slower rate of 0.8 nm/min. As the top \sinn layer is more exposed, it etches slightly faster than the bottom \sinn layer. After the designed release time, the resulting structure has two suspended \sinn layers with nearly identical thickness.

We experimentally demonstrate a high mechanical quality factor of $10^4$ at 102.3 MHz, close to the material limited loss of \sinn at this frequency range \cite{Verbridge:2006ix}. This is more than a threefold improvement over previously demonstrated devices with uncompensated films whose typical measured mechanical quality factors are $Q_m = 2500$~\cite{Zhang:2012ks}. We measure the mechanical quality factor of our devices by coupling a low power continuous wave laser to the devices through a tapered optical fiber, as show in the schematic of Fig. 3a. The devices are characterized inside a vacuum chamber (5-10~mTorr) at room temperature to minimize air damping. The mechanical spectrum can be observed through the optical transmission detected by a fast photodiode (Newport 1811A) which is connected to a radio-frequency spectrum analyzer. We test the optomechanical resonator by tuning the laser to an over-coupled optical resonance near 1530 nm with a loaded optical quality factor of $1.5\times 10^5$. When the low power laser is slightly detuned from the cavity resonance, the thermal Brownian motion of the mechanical resonator is transduced to the optical signal as amplitude modulated radio-frequency (RF) signals. A typical RF spectrum of the detected photocurrent, which is proportional to thermal Brownian mechanical spectrum power density, is shown in Fig. 3c for an optimized cavity. The quality factor $Q_m$, is obtained from a Lorenztian fit through the relation $Q_m = \omega_m/\delta\omega_m$, where $\delta\omega_m = 2\gamma_m$ is the full width half maximum of the thermal Brownian peak and $\omega_m$ and $\gamma_m$ are the mechanical frequency and damping rate. We used an input optical power of 6 $\mu$W, well below the estimated threshold power of regenerative oscillation of 180 $\mu$W. At this input power level, the optomechanical feedback \cite{Kippenberg:2008jva} does not affect the measured $Q_m$ significantly. This is ensured by optimally detuning the laser on both sides of the optical resonance and verify that the difference between the blue and red-detuned Q-measurement is less than 1 percent. The measured mechanical frequencies (dissipation)) are shown as circles in Fig. 3d (Fig. 3e). The mechanical quality factors of the devices with pre-compensated layers have an average mechanical quality factor of $(8.0\pm 0.8)\times 10^3$. 

\begin{figure}[ht]
\begin{center}
\includegraphics[scale=1.1]{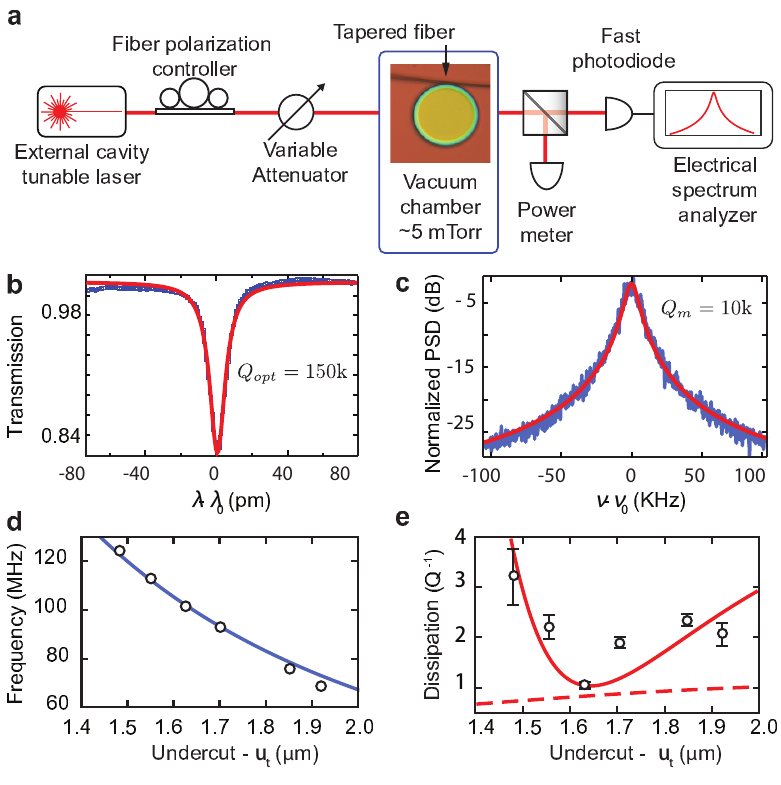}
\caption{\label{fig:experiment} \textbf{Experimental results}.(a) Simplied schematic of the experimental setup. (b) Optical transmission showing a resonance centered at $\lambda_0=$1530.6 nm with a loaded optical quality factor $Q_\text{opt}=1.5\times10^5$. (c) Radio-frequency power spectral density  of the transmitted optical signal. A typical AS mechanical mode resonant frequency centered at $\nu_0=102.3$ MHz for the optimized thickness device showing a quality factor $Q_\text{m}=10^4$. (d,e) Measured mechanical frequency (d) and dissipation (e) of the devices etched through different times. The solid curves are the fitted analytical model prediction, the dashed line is the thermoelastic damping contribution. The error bars in (e) are obtained from the standard deviation among five identical devices}
\end{center}
\end{figure}
We show that the results from the numerical simulations and the experiment can be explained by a simple analytical lumped model of coupled resonators. We decompose the structure into the two freestanding edge resonators and the pedestal resonators as our qualitative analysis described previously. This analytical model agrees with the frequency dependence and the mechanical quality factors observed in both our numerical simulations and experimental results (Fig. 3d,e). In the analysis, we associate a mass-spring lumped model with each resonator identified in Fig. 1e. The resulting coupled system is illustrated in Fig. 1f. Note that when the masses move in opposite phase, there is no net motion of the pedestal and therefore the damping contribution from the pedestal damping $\gamma_p$ is negligible. The normal modes of the coupled system are given by the eigenvectors of the matrix of the system:
\begin{equation}
M(\Omega)=\left(
\begin{array}{ccc}
 i \Delta_p+\gamma_p & i \kappa/2  & i \kappa/2  \\
 i \kappa/2  & i \Delta_t +\gamma_t& i \beta/2 \\
 i \kappa/2  & i \beta/2 & i \Delta_b+\gamma_b \\
\end{array}
\right),
\label{eq:matrix}
\end{equation}
where $\Delta_{p,t,b}\equiv\Omega -\omega_{p,t,b}$ is the detuning of the sought eigenvalue ($\Omega$) and the lumped resonators frequencies ($\omega_{p,t,b}$),  $\kappa$ is the coupling rate between the top and bottom resonators to the pedestal, and $\omega_p$ the pedestal frequency.  $\beta$ is the coupling between the top and bottom resonators and $\gamma_{t,b,p}$ are the damping rate of the three oscillators   We assume the frequency dependence of the freestanding edges ($\omega_t,b$) on the undercut ($u_t$) and thickness ($t_i$) is given by a circular-plate analytical model, $\omega_{i}(t_i)=x t_i u_t^{-2}E^{1/2} [12\rho(1-\nu^2)]^{-1/2}$, where $(E,\rho,\nu)=(250\text{ GPa},3100\text{ kg/m}^3,0.25)$ are respectively the Young modulus, density and the Poisson ratio for \sinn. The numerical factor  $x \sim 1.4$ is comparable to the value obtained by solving the plate problem with a clamped-free boundary condition~\cite{Sun:2009ef}. By solving the characteristic equation given by det$[M(\Omega)] = 0$ we obtain a  complex eigenvalue ($\Omega$) whose real and imaginary parts correspond to the mechanical frequency and damping of the normal modes, respectively. The solid blue and red lines on Figs. 3d and 3e show the fitted model prediction for the mechanical frequency and dissipation ($Q^{-1}=2 \text{Im}\left[\Omega\right]/\text{Re}\left[\Omega\right]$), respectively. The bare frequenies $\omega_{p,t,b} (u_t)$ are calculated from the analytical circular-plate analytical model and is also used to calculated the bare damping rates ($\gamma_t,b =\omega_{p,t,b} /(2Q_{t,b})$), with the bare quality factors ($Q_{t,b}$) inferred from FEM simulations. Since the model parameters impact very distinctively the real and imaginary parts of the complex eigenvalue, they were iteratively adjusted using both the measured frequencies (Fig. 3d) and damping rates (Fig. 3e). The fitted parameters are given by $(Q_p, \kappa, \beta,x) = (1.2; 110 \text{Mhz}, 5.96 \text{Mhz}, 1.36)$ for which their initial values are estimated also by FEM simulations. When the structural loss is eliminated, the dominant loss will be thermoelastic damping. We show in Fig. 3e,  dashed-red line, the fundamentally limited dissipation based on thermoelastic damping  prediction using the typical \sinn parameters $(c_p, \kappa_t, \alpha, T) = (710~\text{JK}^{-1}; 3.2~\text{Wm}^{-1}\text{K}^{-1}; 2\times 10^{-6}~\text{K}^{-1}, 300~\text{K})$, representing respectively the specific heat, thermal conductivity, thermal expansion coefficient, and temperature~\cite{Lifshitz:2000dj,Verbridge:2006ix,Sun:2010ie,Yasumura:cv}. This shows that our demonstrated device is within a factor of 1.3 of the material limited damping. Despite the simplicity of the model, the obtained fit parameters are in good agreement with values inferred from the FEM simulations.\\

Reducing the structural loss using destructive elastic wave interference is not only limited to the double disk optomechanical oscillator. Using the same methodology, one could design for example an identical pair of loosely spaced singly or doubly clamped cantilevers. When they are excited in an anti-symmetric fashion, their support loss can be eliminated. Our method opens a path towards the deterministic design of micro- and nanomechanical resonators that are limited only by material losses. 

We acknowledge Prof. Paulo Nussenzveig for helpful discussions. This work was performed in part at the Cornell Nano-Scale Science and Technology Facility (a member of the National Nanofabrication Users Network) which is supported by NSF, its users, Cornell University and Industrial users. The	authors	gratefully	acknowledge	support	from DARPA for award W911NF‐11‐
1‐0202 supervised by Dr. Jamil Abo‐Shaeer. G.S.W acknowledges FAPESP and CNPq INCT Fotonicom for financial support in Brazil.

\tocless{\bibliography{highq}}


\end{document}